\documentclass[10pt,aps,pra,twocolumn,showpacs,superscriptaddress,nofootinbib %
 reprint,
]{revtex4-2}

\usepackage{amsmath}
\usepackage{amssymb}
\usepackage{amsfonts}
\usepackage{mathrsfs}
\usepackage{amsthm}
\usepackage{mathtools}
\usepackage{braket}
\usepackage{physics}
\usepackage{nicematrix}
\usepackage{bm} 
\usepackage{dsfont}

\theoremstyle{definition}

\usepackage{graphicx}
\usepackage{dcolumn}
\usepackage{bm}
\usepackage{appendix}
\usepackage{amsfonts}
\usepackage[normalem]{ulem}
\usepackage{xcolor}
\usepackage{booktabs}
\usepackage{comment}
\usepackage{lmodern}
\usepackage{stackrel}
\usepackage{bbold}
\usepackage{soul}
\usepackage{hyperref}

\usepackage[small,bf,justification=Justified]{caption}
\newcommand{\com}[1]{} 

\begin{document}

\title{Expressibility, entangling power and quantum average causal effect for causally indefinite circuits}

\author{Pedro C. Azado}
\email{pedroazado@ifsc.usp.br}
\affiliation{
 Instituto de Física de São Carlos, Universidade de São Paulo, CP 369, 13560-970, São Carlos, SP, Brasil.}
 \affiliation{International Iberian Nanotechnology Laboratory (INL), Braga, Portugal}
 
\author{Guilherme I. Correr}
\affiliation{
 Instituto de Física de São Carlos, Universidade de São Paulo, CP 369, 13560-970, São Carlos, SP, Brasil.}
 \affiliation{QTF Centre of Excellence and Department of Physics, University of Helsinki, FIN-00014 Helsinki, Finland}
 
\author{Alexandre Drinko}
\affiliation{
 Instituto de Física de São Carlos, Universidade de São Paulo, CP 369, 13560-970, São Carlos, SP, Brasil.}
 \affiliation{Faculty of Sciences, UNESP—São Paulo State University, 17033-360 Bauru-SP, Brazil}
 
\author{Ivan Medina}
\affiliation{
 Instituto de Física de São Carlos, Universidade de São Paulo, CP 369, 13560-970, São Carlos, SP, Brasil.}
 \affiliation{School of Physics, Trinity College Dublin, Dublin 2, Ireland.}

\author{Askery Canabarro}
\affiliation{
 Instituto de Física de São Carlos, Universidade de São Paulo, CP 369, 13560-970, São Carlos, SP, Brasil.}
\affiliation{Grupo de Física da Matéria Condensada, Núcleo de Ciências Exatas - NCEx, Campus Arapiraca, Universidade Federal de Alagoas, 57309-005 Arapiraca, Alagoas, Brazil}
\affiliation{Department of Physics, Harvard University, Cambridge, Massachusetts 02138, USA}
\affiliation{Quantum Research Center, Technology Innovation Institute, Abu Dhabi, UAE}

\author{Diogo O. Soares-Pinto}
\email{dosp@ifsc.usp.br}
\affiliation{
  Instituto de Física de São Carlos, Universidade de São Paulo, CP 369, 13560-970, São Carlos, SP, Brasil.}

\begin{abstract}
Parameterized quantum circuits are the core of new technologies such as variational quantum algorithms and quantum machine learning, which makes studying its properties a valuable task. We implement parameterized circuits with definite and indefinite causal order and compare their performance under particular descriptors. One of these is the expressibility,  which measures how uniformly a given quantum circuit can reach the whole Hilbert space. Another property that we focus on this work is the entanglement capability, more specifically the concurrence and the entangling power. We also find the causal relation between the qubits of our system with the quantum average causal effect measure. We have found that indefinite circuits offer expressibility advantages over definite ones while maintaining the level of entanglement generation. Our results also point to the existence of a correlation between the quantum average causal effect and the entangling power.
\end{abstract}

\maketitle

\section{\label{sec:introduction}Introduction}

Claims of quantum advantage are recurrent~\cite{gibney2019hello,arute2019quantum}, but not without controversies~\cite{pan2022solving}. Despite adversities, noisy intermediate-scale quantum (NISQ) computers might be useful regardless~\cite{preskill2018quantum}. These devices, with qubit counts ranging from tens to hundreds, offer the potential to solve complex problems beyond the reach of classical computers. However, they remain constrained by high noise levels, operational errors, limited qubit counts, and restricted interconnectivity, all of which pose significant challenges when working with NISQ computers.

One of the key strategies employed in harnessing the potential of NISQ computers is the utilization of parameterized quantum circuits (PQCs). These circuits are designed with adjustable parameters, allowing for a more extensive exploration of the Hilbert space, and by combining them with a classical optimizer, we get what are called variational quantum algorithms (VQAs)~\cite{Cerezo2021VqaReview}, which turned out to be one of the prominent strategies to obtain quantum advantage so far.

This better exploration of the Hilbert space, provided by PQCs, is quantified by the concept of \textit{expressibility}~\cite{sim2019expressibility}, which measures how uniformly a quantum circuit can access different states within the Hilbert space. Expressibility is a crucial metric for assessing the capabilities of NISQ computers in performing quantum computations since the lack thereof could lead to circuits that cannot reach the desired state, or, on the other hand, too much expressibility could lead to optimization problems such as barren plateaus~\cite{larocca2024review}.

Another critical aspect is the evaluation of circuit entanglement, often quantified using measures such as the concurrence~\cite{hill1997entanglement}, entangling power~\cite{zanardi2000entangling,eisert2021entangling} and many more~\cite{guhne2009entanglement}. Understanding the entanglement capabilities of PQCs is vital for quantum algorithms and their applications~\cite{ortiz2021entanglement,yao2024avoiding}, since it might constitute a resource to obtain quantum advantage.

Recently, other interesting resource, the indefinite causal order of operations, has drawn the attention of the quantum computing community. It has been found that by applying an indefinite causal order of operations 
\cite{oreshkov2012quantum,brukner2014quantum} in the circuit, one can achieve an advantage over causally ordered circuits by means of reducing the computational query complexity~\cite{procopio2015experimental,liu2023experimentally,araujo2014computational}.  This indefinite order  can be achieved, for example, through a specific operation called the  quantum switch~\cite{chiribella2013quantum}, which allows for the superposition of quantum operations.
Along with the quantum switch, it is also possible to implement the quantum time flip~\cite{chiribella2022quantum,stromberg2024experimental}, which belongs to a more general class of indefinite processes, called indefinite time direction~\cite{chiribella2022quantum}. Investigating the relationship between indefinite causal structures and improved utilization of the Hilbert space in parameterized quantum circuits could be a valuable endeavor to make the most out of NISQ computers.

This raises the question: is indefiniteness a measurable property, or is it binary? A productive approach could be to examine causality as understood in classical physics~\cite{pearl2009causality} and adapt these concepts to the quantum domain. This is precisely what has been done in Ref.~\cite{hutter2023quantifying}, where the quantum Average Causal Effect (qACE) was defined and later tested experimentally in Ref.~\cite{agresti2022experimental}. Although not the only possible way to measure this indefiniteness~\cite{goswami2024maximum}, the qACE will be the approach explored in this work.

The primary objective of this study is to investigate and compare the properties of quantum circuits with both definite and indefinite causal structures, focusing on expressibility, entanglement capability, and the quantum average causal effect (qACE). Specifically, we aim to determine if and how causally indefinite circuits, such as those utilizing the quantum switch and time flip operations, exhibit advantages over causally definite circuits. By examining expressibility, which measures the circuit's ability to uniformly explore the Hilbert space, and entangling power, which quantifies the circuit's capacity to generate entanglement, we seek to evaluate the potential of indefinite causal structures as a resource for enhancing quantum algorithms on near-term noisy quantum devices. Additionally, we analyze the quantum average causal effect to quantify the causal influence between qubits within these circuits, such that a relation between these quantifiers might be drawn.

The structure of the paper is as follows: In Sec. II, we introduce the causally definite and indefinite quantum circuits that will be compared using the quantifiers outlined in Sec. III, which include expressibility, concurrence, entangling power, and qACE. Sec. IV presents a discussion of our results, followed by Sec. V, where we conclude with our insights on the topic.

\section{\label{sec:QC}Quantum Circuits}

\subsection{\label{subsec:qSwitch}Quantum switch}

\begin{figure}[t]
	\centering
	\includegraphics[width=\columnwidth]{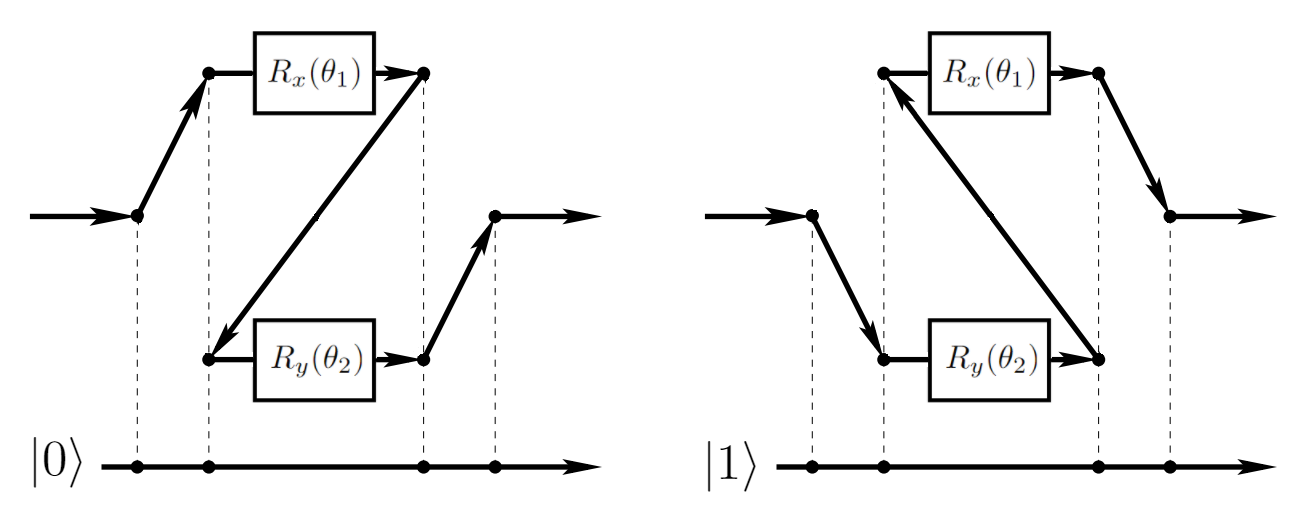}
	\caption{Graphic illustration of the operations that compose the quantum switch. 
The order of these operations $R_x(\theta_1)$ and $R_y(\theta_2)$ are dependent on the state of the control qubit. When applied in a superposition of orders, this results in an indefinite causal structure, meaning the order of the operations is not predetermined, but exists in a quantum superposition. Therefore, for our case, the initial state of the control qubit is prepared in the state $\ket{+}=(\ket{0}+\ket{1})\sqrt{2}$.}
	\label{switches}
\end{figure}

The first indefinite circuit we will deal with is composed solely of the quantum switch~\cite{chiribella2013quantum}. It is a supermap that takes two operations as input and returns another as output, and it can be divided into two quantum systems: The control and the target. The control is a qubit that will be prepared in a superposition state, such that the order of operations that will be acted upon the target system is controlled by the state of the control. It would be problematic to illustrate the quantum switch in the usual quantum circuit diagram, where the direction of time and order of operations is well defined. Then, in Fig.~\ref{switches} we have the usual approach to portray this supermap with the operations that we will be using in this work.

The quantum switch can be written as a unitary gate that acts on a quantum circuit as
\begin{equation}\label{switcheq}
    \mathcal{S} = \ketbra{0}{0} \otimes U_1 U_2 + \ketbra{1}{1} \otimes U_2 U_1,
\end{equation}
where the control qubit is prepared in a $\ket{+}=(\ket{0}+\ket{1})/\sqrt{2}$ state and it gets entangled with the different orders for the operations $U_1$ and $U_2$.

More specifically for our case, we are interested in seeing how the quantum switch behaves with parameterized gates as operations, so we implement it with rotations in the $x$ and $y$ axes such that the operator reads
\begin{eqnarray}
    \mathcal{S}(R_x,R_y) = &\ketbra{0}{0}& \otimes R_x(\theta_1)R_y(\theta_2) \nonumber \\ 
   + &\ketbra{1}{1}& \otimes R_y(\theta_2)R_x(\theta_1).
\end{eqnarray}
The rotations are chosen based on the initial state of the target qubit, which is assumed to be $\ket{0}$. While this is not the only possible arrangement of rotations and parameters, we focused on the simplest case to ensure consistency. The number of parameters was kept fixed across all circuits to allow for a fair comparison between them.


Even though employing the quantum switch might be feasible, one could still ask if it is advantageous, which is what has been investigated by Refs.~\cite{renner2022computational,araujo2017quantum} with positive results. Not only that, it has been proven useful in quantum metrology~\cite{yin2023experimental,zhao2020quantum}, parameter estimation~\cite{procopio2023parameter}, quantum communication~\cite{guerin2016exponential,ebler2018enhanced,chiribella2021indefinite}, thermometry~\cite{mukhopadhyay2018superposition} and many others~\cite{mukhopadhyay2020superposition,loizeau2020channel,caleffi2020quantum,procopio2020sending,felce2020quantum,simonov2022work,cao2022quantum,simonov2023universal}. Just as important, the authors of Ref.~\cite{capela2022indefinite} showed that indefinite causal order is not always an advantageous thermodynamic resource.

\subsection{\label{subsec:ITD}Quantum time flip}

The other class of indefinite circuits we will be dealing with is one composed by an operation called time flip~\cite{chiribella2022quantum,stromberg2024experimental}. Just as the quantum switch is the supermap that introduces an indefinite causal order in the quantum circuit, the time flip operation characterizes the indefinite time direction process~\cite{chiribella2022quantum}. This class of processes takes advantage of time-symmetric operations to coherently superpose its forward and backward time-direction operations, analogous to the quantum switch when it superposes the order of two operations.

Since we are dealing with two-dimensional unitary operations, both the inverse and the transpose can be regarded as the time reversal operation. For higher dimensions, however, the adjoint fails to generate completely positive maps when applied locally to bipartite systems~\cite{chiribella2008transforming,chiribella2009theoretical,bisio2019theoretical}. Therefore, the time flip operation is given by
\begin{equation}
    \mbox{TF}(U) = \ketbra{0}{0} \otimes U + \ketbra{1}{1} \otimes U^{T},
\end{equation}
where the control qubit is prepared in a superposition and the transpose is chosen to be the backward in time operation. However, if we take $U = R_x(\theta_1)R_y(\theta_2)$, we would end up with $\mbox{TF} = \ketbra{0}{0} \otimes R_xR_y + \ketbra{1}{1} \otimes R_y^TR_x^T$, which is the same as the quantum switch for these two rotation operators, since $R_x^T = R_x$ and $R_y^T(\theta) = R_y(-\theta)$. Then, the time flip operator we shall be using reads
\begin{eqnarray}
    \mbox{TF}(R_x,R_y) = &\ketbra{0}{0}& \otimes R_x(\theta_1)R_y^T(\theta_2) \nonumber \\ 
    + &\ketbra{1}{1}& \otimes  R_x^T(\theta_1) R_y(\theta_2),
\end{eqnarray}
which is the one from the quantum game in Ref.~\cite{stromberg2024experimental}, used as an example to show that the time flip operation can be advantageous in quantum information protocols. 

Essentially, the quantum game describes the situation where two black boxes implement two unitary operations $U$ and $V$ that either satisfy the relation $UV^T=U^TV$ or $UV^T=-U^TV$. The goal of the game is to find out which of these relations is true. By using a process with indefinite time direction, one can show that the player can win the game with certainty, which is not possible even with the use of the quantum switch. This game was used to experimentally certify the advantage of the time flip operation over operations with definite time and indefinite causal order in~\cite{stromberg2024experimental}.

\subsection{Comparison to causally ordered circuits}

Lastly, we propose some circuits to serve as comparisons to those with indefinite structures. To choose suitable comparisons, we were looking for circuits that, most importantly, had a definite causal or temporal structure, circuits that could resemble the operations that we were doing with the quantum switch or the time flip, and circuits commonly used in variational algorithms. All the while keeping the number of parameters and qubits the same throughout.

There is the case that an arbitrary two-qubit unitary could serve as a general benchmark, however, such a comparison would not align with the goals of this study. First, expressibility for an arbitrary unitary would trivially reach the Haar measure, which is not necessarily desirable for parameterized quantum circuits (PQCs) due to potential optimization challenges such as barren plateaus~\cite{mcclean2018barren}. Second, entangling power and qACE depend on the unitary's structure, and comparing structured PQCs with unstructured unitaries would obscure the role of causal indefiniteness. Additionally, an arbitrary two-qubit gate requires at least three parameters, whereas our study maintains a fixed number of parameters across all circuits to ensure a fair comparison. Finally, unstructured unitaries are less practical for variational algorithms, which typically rely on structured, hardware-efficient ansätze. By focusing on these four circuits, we isolate the impact of causal indefiniteness while maintaining relevance to near-term quantum computing applications.

\begin{figure}[t]
	\centering
	\includegraphics[width=\columnwidth]{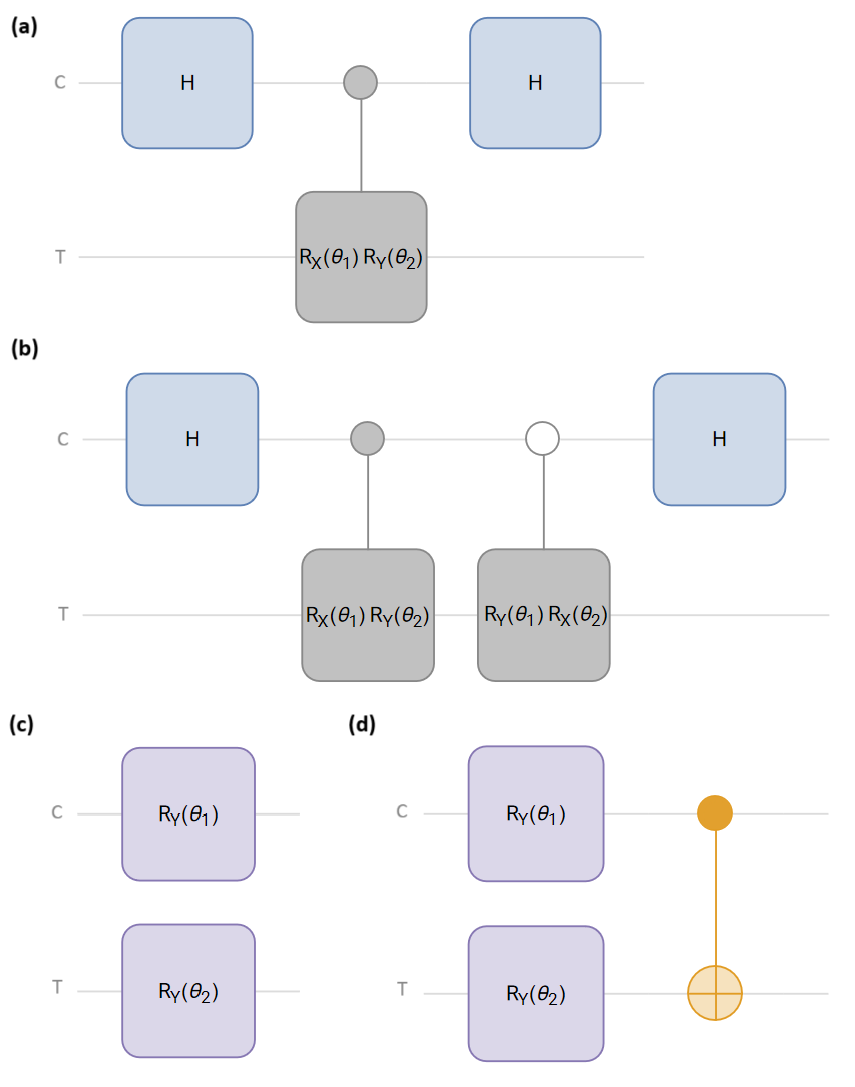}
	\caption{Definite structure circuits for comparison with the indefinite ones. $C$ represents the control qubit and $T$ the target of the operations.
 (a) A quantum circuit where the control qubit determines the order of operations $R_x(\theta_1)$ and $R_y(\theta_2)$ based on the state of the control, we will call it $R_{xy}$.
 (b) A similar circuit where both orders of operations are applied on the target qubit, depending on the control qubit's state, simulating the action of the quantum switch with more operations. We will refer to it as $R_{xy}R_{yx}$.
 (c) The separable circuit (SEP), where the rotation operations are applied independently on each qubit without any entangling gates.
 (d) The hardware-efficient ansatz (HEA) is a quantum circuit with usual entangling operations, here shown with a CNOT, usually designed to minimize circuit depth in variational algorithms.}
	\label{circuitos}
\end{figure}

In the first circuit, Fig.~\ref{circuitos}a, we employ a combination of Hadamard gates and parameterized $R_x(\theta)$ and $R_y(\theta)$ rotations. This circuit is particularly relevant because it represents one of the operations performed by the quantum switch but without the superposition of causal orders. By eliminating the superposition, we ensure a definite causal structure while retaining a similar gate architecture. The Hadamard gates will create the necessary superpositions for the controlled gates, while the parameterized rotations enable flexible control over the circuit’s expressibility and entangling capabilities.

The second circuit, Fig.~\ref{circuitos}b, is based on a simulation of the quantum switch~\cite{fellous2023comparing}, where we have two controlled operations acting on the qubits with reversed orders of application, specifically $R_x(\theta_1) R_y(\theta_2)$ and $R_y(\theta_1) R_x(\theta_2)$. This allows us to observe how expressibility and entanglement capability is affected when the quantum switch’s indefinite causal order is replaced by a definite one. One important remark is that this circuit is considered a simulation of the quantum switch only when we trace out the control qubit, which means that its output with two qubits is not the same as the one from the quantum switch itself, leading to different measures of expressibility and entanglement capability in the following sections. There are different ways to simulate the quantum switch~\cite{bavaresco2024can} and we chose this one in specific for its limited number of qubits and easier conceptualization.

The third circuit, Fig.~\ref{circuitos}c, known as the separable circuit (SEP), contains a layer of parameterized $R_y(\theta)$ rotations on each qubit. This design introduces no entanglement between the qubits, which makes it useful as a baseline for comparing the influence of entanglement on expressibility and optimization performance. In the context of variational quantum algorithms, such circuits are often used to probe states without entanglement~\cite{drinko2024benchmarking}.

Finally, the fourth circuit, Fig.~\ref{circuitos}d, which could be classified as a hardware-efficient ansatz (HEA), introduces a CNOT gate alongside the $R_y(\theta)$ rotations. This type of circuit is commonly used in parameterized quantum algorithms due to its balance of expressibility and feasibility on current quantum hardware, though usually with more layers. The CNOT gate acts as an entangling operation, but unlike in the quantum switch, this entanglement follows a fixed, predefined temporal order. This design makes it ideal for studying how a circuit with a fixed causal structure compares against one with indefinite causal order while keeping the gate count and number of parameters constant.

By incorporating these circuits into our comparison, we can directly evaluate how different types of causal structures, ranging from definite to indefinite, affect a circuit’s ability to generate entanglement, its expressibility, and the causal influence between qubits. Maintaining consistency in the number of parameters and qubits across all circuits ensures that differences in performance are tied to the causal structure rather than to circuit complexity.

For future discussion of the results, we will use $R_{xy}$ to represent the controlled rotations, that is
\begin{equation}
    R_{xy}(\theta_1,\theta_2) = \ketbra{0}{0} \otimes \mathbb{1} + \ketbra{1}{1} \otimes R_x(\theta_1)R_y(\theta_2),
\end{equation}
and also use as a label to the first and second circuits, as $R_{xy}$ (Fig.~\ref{circuitos}a) and $R_{xy}R_{yx}$ (Fig.~\ref{circuitos}b)  respectively.

\section{\label{sec:quantifiers}Quantifiers}

\subsection{\label{subsec:expressibility}Expressibility}

First proposed in Ref.~\cite{sim2019expressibility}, expressibility has been used to investigate the occurrence of barren plateaus~\cite{holmes2022connecting} and to analyze a common structure for PQCs, that is, one of alternating layered ansatz~\cite{nakaji2021expressibility}. The idea has also been present to better understand group theoretic characteristics of quantum circuits~\cite{ragone2024lie,diaz2023showcasing,larocca2022group,ragone2022representation}.

Simply put, we can define expressibility as the ability of a given parameterized quantum circuit to generate (pure) states that are a good representation of the whole Hilbert space. In other words, we want to know how uniformly distributed in the Hilbert space are the states generated by a given ensemble of unitaries.

To find how uniformly distributed the outcome of a given ensemble is, we compare the distribution of states generated by a given PQC and the uniform distribution of states, namely, the ensemble of Haar-random states~\cite{mezzadri2006generate,mele2024introduction}
\begin{equation}
    A^{(t)}=\int_{Haar} (\ketbra{\psi}{\psi})^{\otimes t} d\psi - \int_{\Theta} (\ketbra{\phi_\theta}{\phi_\theta})^{\otimes t} d\theta,
\end{equation}
where the first integral is over the states distributed with respect to the Haar measure and the second is the distribution of states generated by sampling different values for the parameters in the PQC. The $(\cdot)^{\otimes t}$ means that this is a comparison over a $t$-design, i.e., up until the $t$-th moment of such distributions~\cite{unitary_designs, polynomial_t-designs1, polynomial_t-designs2}. It has been demonstrated that this relation leads to a comparison of distributions considering fidelities between states generated by the circuit and the Haar random states~\cite{sim2019expressibility}.
We then compare these distributions with the relative entropy, or Kullback-Leibler divergence~\cite{kullback1951information}. The expressibility is then given by
\begin{equation}
    \text{Expr}:=D_{KL}\left(P_{PQC}(F(\Theta))||P_{\text{Haar}}(F)\right),
\end{equation}
where the distribution generated by our PQCs depends on the parameter space $\Theta$ and on the fidelities $F(\theta,\phi):=|\braket{\psi(\theta)}{\psi(\phi)}|^2$ among final states of the same quantum circuit initialized with different parameters $\theta$ and $\phi$. In addition, the relative entropy $D_{KL}$ is given by
\begin{equation}
    D_{KL}(P||Q) = \sum_x P(x) \log\left({\frac{P(x)}{Q(x)}}\right).
\end{equation}
By using the KL divergence we have an operationally meaningful measure of expressibility, where the closer our result is to zero, the closer our PQC will be from generating random states.

To compute the expressibility, we first act upon the initial state $\ket{00}$ a chosen circuit with uniformly sampled parameters to find a final state $\ket{\psi(\theta)}$. We then sample the parameters one more time and compute the fidelity $F(\theta,\phi):=|\braket{\psi(\theta)}{\psi(\phi)}|^2$ between the final states. These fidelity values are used to construct a histogram that will be compared with the analytically found Haar histogram using the relative entropy $D_{KL}$. We end up with a number that characterizes the expressibility and it has no dependence on the chosen parameters since, by uniformly sampling them, we are effectively averaging them out.

\subsection{\label{subsec:concurrence}Concurrence}

We have also decided to include entanglement measures and given that we have a simple circuit with only two qubits, the concurrence~\cite{hill1997entanglement} seemed especially appropriate. For pure states, it reads
\begin{equation}
    C(\ket{\psi}) = \sqrt{2[1-\mbox{Tr}(\rho_B^2)]},
\end{equation} 
where $\rho_B$ is the reduced density matrix $\mbox{Tr}_A(\rho_{AB})$. We consider $A$ as the control qubit and $B$ as the target qubit.

However, even though we take the average of the concurrence over all parameters in the parameterized circuit, the concurrence will be dependent on the initial state on the qubits. One way to circumvent this issue is to take an average of all possible initial product states and better evaluate how the unitary itself is generating entanglement. With this in mind, we will also work with the entangling power~\cite{zanardi2000entangling}.

\subsection{\label{subsec:entpower}Entangling Power}

As defined in Ref.~\cite{zanardi2000entangling}, the entangling power is a entanglement measure that does not depend on the initial state. It is defined by
\begin{equation}
    e_p(U) = \overline{E(U \ket{\psi_1} \otimes \ket{\psi_2})\!}^{(\psi_1,\psi_2)},
\end{equation}
where the bar represents the average over all initial product states $\ket{\psi_1} \otimes \ket{\psi_2} \in \mathcal{H}_1 \otimes \mathcal{H}_2$, according to a probability distribution $p(\psi_1,\psi_2)$ and with dimensions $d_1$ and $d_2$, respectively. $E(\cdot)$ measures the entanglement by means of the linear entropy,
\begin{equation}
    E(\ket{\psi}) = 1 - \mbox{Tr}_1(\rho^2),
\end{equation}
where $\rho = \mbox{Tr}_2(\ketbra{\psi}{\psi})$.

This measure is particularly useful for its sole dependence on the applied unitary, which is precisely what we look for when analyzing causally indefinite unitaries, such as the quantum switch and the indefinite time direction process.

When the average is taken over the uniform distribution of product states~\cite{zanardi2000entangling}, the entangling power is given by
\begin{eqnarray}
    e_{p_0}(U) &=& 1 - C_{d_1} C_{d_2} \sum_{\alpha=0,1} I_{\alpha}(U), \\
    I_{\alpha}(U) &=& t(\alpha) + \expval{U^{\otimes 2} (T_{1+\alpha,3+\alpha}) U^{\dagger \otimes 2}, T_{13}},
\end{eqnarray}
where $t(\alpha)= \mbox{tr}(T_{1+\alpha,3+\alpha})$ and $C_d^{-1}=d(d+1)$. Since we have the simple case of two qubits, we have that $d_1 = d_2 = d = 2$, which further simplifies these equations.  In addition, $T_{ab}$ represents the SWAP operator in $d\times d$ dimensions and the subscripts are the respective bipartitions to be swapped.

\subsection{\label{subsec:qACE}qACE}

When dealing with causally indefinite circuits, such as the quantum switch and the indefinite time direction process, a fundamental question arises: how causally connected are the subsystems within these channels? To address this, we employ the quantum average causal effect (qACE), a quantum adaptation of a well-established causal influence quantifier~\cite{hutter2023quantifying,agresti2022experimental}. Similarly, recent proposals have introduced measures to quantify the strength of causal relationships in these systems~\cite{goswami2024maximum}.

This quantifier measures the change in a physical system induced by changes in another system under the action of a given unitary operation. For our two-qubit case specifically, we are interested to know how the target qubit changes as we perform a do-intervention on the control qubit, i.e., how the state of the target qubit is affected by forcefully setting the control qubit to a specific pure state. Since the choice for this intervention and for the input for the target qubit are as good as any, we take the average over the uniform (Haar) distribution~\cite{mezzadri2006generate,mele2024introduction} over both. Finally, after two such interventional final states are realized, we compare them with the trace distance
\begin{eqnarray}
    TD(\rho,\sigma) = \frac{1}{2} || \rho - \sigma ||_1 = \frac{1}{2} \mbox{Tr} \Big( \sqrt{(\rho - \sigma)^2} \Big),
\end{eqnarray}
with $(\rho - \sigma)^2 = (\rho - \sigma)^\dagger(\rho - \sigma)$.

The whole process can be encapsulated by the following equation
\begin{equation}
    ACE_Q(U) = \mathop{\mathds{E}}_{\ket{a}} \mathop{\mathds{E}}_{\ket{b}} TD[\rho(b|do(a)),\rho(b|do(a^\perp)],
\end{equation}
where $\mathds{E}$ represents the expected value over the initial states $\ket{a}$ and $\ket{b}$, $\ket{a^\perp}$ is the antipodal state from $\ket{a}$ and
\begin{eqnarray}
    \rho(b|do(a)) = \mbox{Tr}_A \Big( U \big( \ketbra{a}{a} \otimes \ketbra{b}{b} \big) U^{\dagger} \Big),
\end{eqnarray}
represents the density matrix for the state $b$ given the intervention $do(a)$. Using the quantum switch as an example, it is clear that, given an arbitrary state $\ket{a}$ for the control, the maximum influence over the target qubit comes from choosing either $\ket{a}$ or $\ket{a^\perp}$ as intervention.

\section{\label{sec:results}Results}

\subsection{Expressibility}

\begin{figure}[t]
	\centering
	\includegraphics[width=\columnwidth]{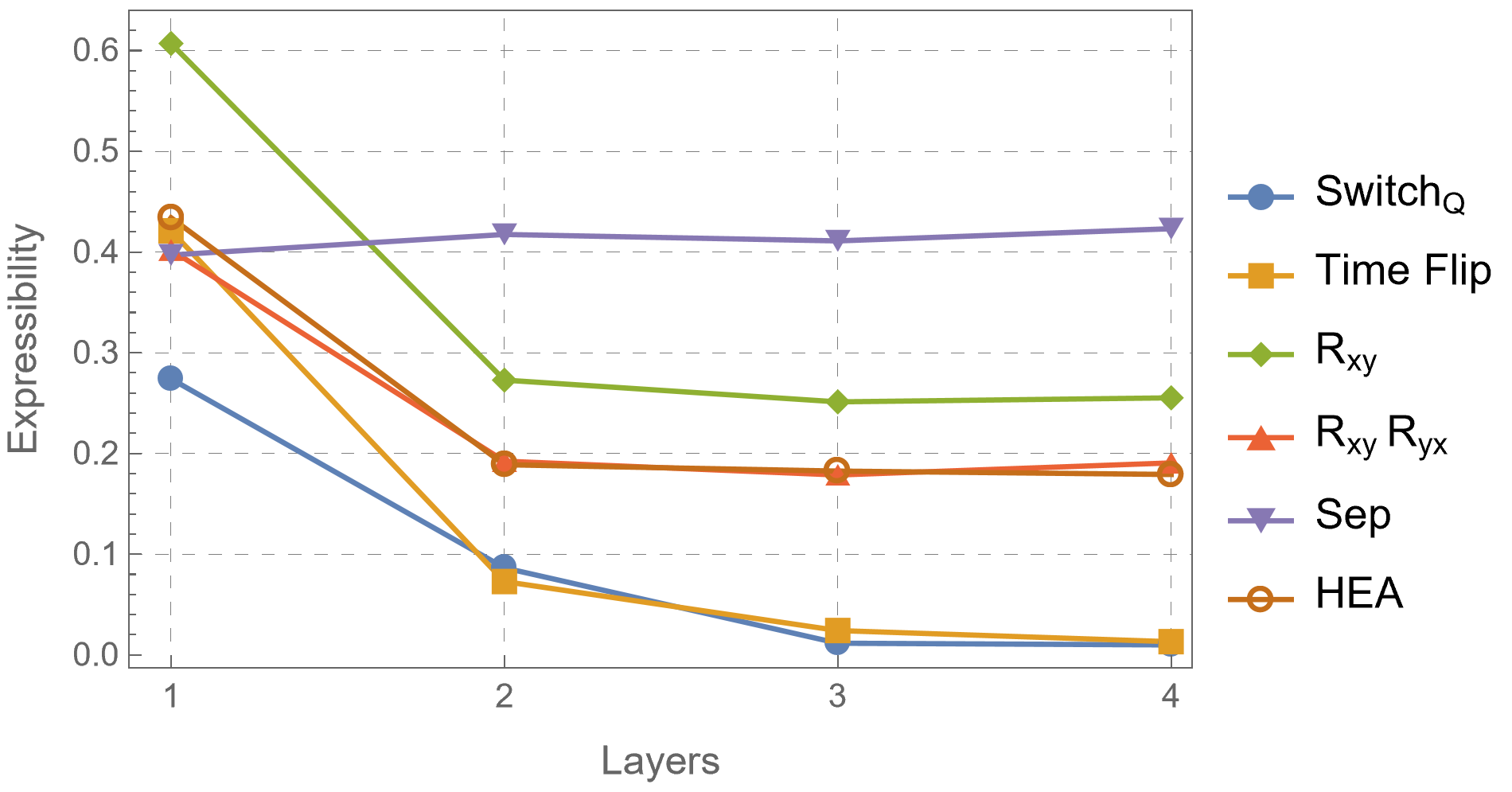}
	\caption{Expressibility by number of layers, with each curve representing a different circuit. Each layer is a repetition of the circuits defined in Sec.~\ref{sec:QC} with the number of parameters following the number of new applied rotations. Here, expressibility denotes the relative entropy, $D_{KL}$, which presents lower values for more expressible circuits, i.e., generating states more uniformly distributed, as defined in~\cite{sim2019expressibility}. From this figure, we can see that expressibility is higher (i.e., lower $D_{KL}$ values) in the chosen indefinite circuits compared to the definite ones, with the Quantum Switch and the Time Flip reaching these lower values more rapidly as the number of layers increases.}
	\label{expressibility}
\end{figure}

The expressibility measure used in this analysis is inversely related to the circuit’s capacity to explore the Hilbert space, where lower values indicate higher expressibility. Fig.~\ref{expressibility} illustrates the expressibility of several quantum circuits, comparing indefinite causal and temporal structures, such as the quantum switch and time flip operation, with standard definite circuits and how they evolve with each added layer. Whereas by layer, we mean repetitions of the previously defined circuits with the number of parameters growing with the number of new rotations applied. As expected, across all configurations except for the separable circuit, the expressibility decreases as the number of layers increases. This indicates that deeper circuits become more effective at uniformly covering the Hilbert space.

At first glance, one interesting aspect of the separable circuit curve is that it is the only one that remains relatively constant as more layers are combined. It is easy to see why this would happen when we consider that the qubits in this circuit are not entangled with each other, which means that the state of spaces is separable and that by adding the same rotation again with each layer, we are still exploring the same segment of the Bloch sphere. For a single layer, however, we see that it has an expressibility comparable to the $R_{xy}R_{yx}$, HEA and the one with the time flip operation, even though these already generate entanglement in the first layer. This could be explained by the assumption that a single entangling layer is still not enough to saturate the entanglement in a two-qubit circuit, which also explains why all of them have better expressibility for more layers.

Now to the indefinite circuits, we have the quantum switch and the time flip expressibility curves. While for a single layer the time flip curve seems to be on par with the rest of the compared circuits, it has a steeper increase in expressibility as the number of layers rises, reaching the quantum switch after the second is added. As for the quantum switch itself, we have found that it has a better expressibility even for lower-depth circuits than the causally definite ones.

From the analysis, we observe that the quantum switch exhibits superior expressibility compared to the causally definite circuits, such as the HEA and the standard rotation-based circuits. While the time flip circuit shows strong expressibility at greater depths, it only starts to approach the quantum switch's performance after two layers. This suggests that indefinite causal structures like the quantum switch and time flip offer an enhanced ability to explore the state space, particularly as the number of layers increases. In contrast, definite circuits with entanglement, such as the $R_{xy}R_{yx}$, exhibit slower improvement in expressibility with increasing layers, while the separable circuit, which lacks entanglement, remains less effective in state exploration. 

Nevertheless, one must take into account that those are limited systems with a limited number of rotations and parameters. Depending on the goal, reaching higher expressibility could be undesired, or perhaps, simply adding more parameters and circuit depth might be easier than implementing indefinite circuits, since we can see that already with two layers, definite circuits have similar expressibility than the quantum switch for a single one.

\subsection{Entanglement}

\begin{figure*}[t]
	\centering
	\includegraphics[width=\textwidth]{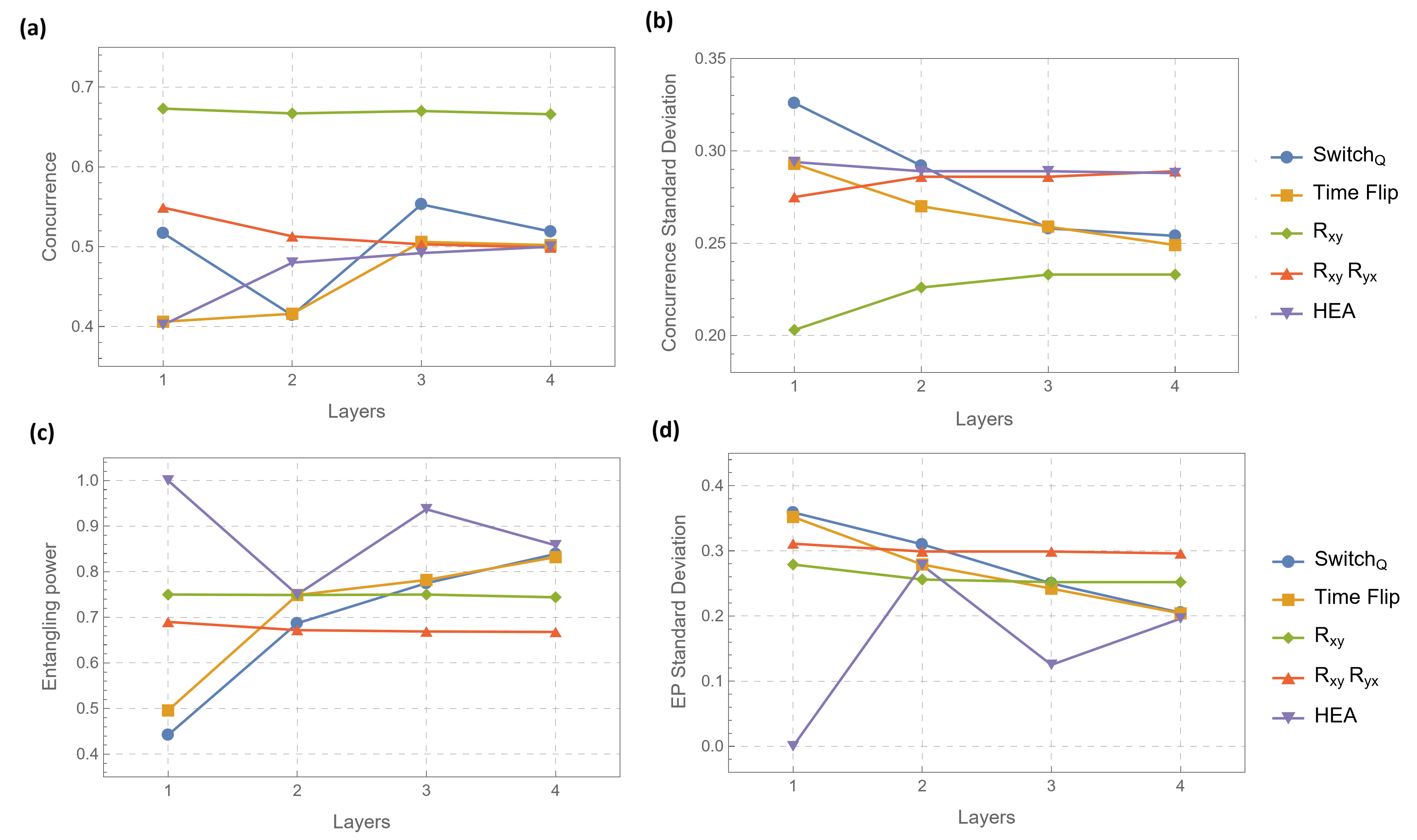}
	\caption{Entanglement results between definite and indefinite circuits. In (a) and (b) we have the concurrence and its standard deviation, whereas in (c) and (d) we have the entangling power and its standard deviation, all by the number of layers. Each curve represents a different circuit from Sec.~\ref{sec:QC} that can generate entanglement. From panels (a) and (b), we observe that the concurrence and its standard deviation for the indefinite circuits are comparable to—or lower than—those of the definite circuits, converging to similar values as the number of layers increases. In contrast, the entangling power shown in (c) and (d) reveals a more distinct behavior: indefinite circuits initially generate lower entangling power with higher standard deviation at fewer layers, but as the number of layers increases, their entangling power surpasses that of the definite circuits, while the standard deviation decreases.}
	\label{entanglement}
\end{figure*}

We have our results for the entanglement analysis in Fig.~\ref{entanglement}a and Fig.~\ref{entanglement}c, where we have both the concurrence and the entangling power, together with their standard deviations, in Fig.~\ref{entanglement}b and Fig.~\ref{entanglement}d, respectively. Compared to the expressibility results, we have taken out the separable circuit curve from the figures, since it does not generate entangled states, and the system we consider encompasses both the control and the target qubits. It is also important to note that to plot these entanglement values we have taken the average over the rotation parameters and found their standard deviations. For the entanglement power specifically, there is the average over the initial product states from the definition and also this average over the parameter space.

The entanglement capabilities of both indefinite and definite quantum circuits, as reflected in the concurrence results, reveal key differences in performance as circuit depth increases. Among the circuits, the $R_{xy}$ consistently demonstrates the highest entanglement measure, with a stable concurrence of around 0.7 across all layers. This stability suggests that this circuit is well-suited for generating and sustaining entanglement, maintaining its high performance even as the number of layers grows. The low standard deviation observed for the $R_{xy}$ circuit further supports this conclusion, indicating that the entanglement produced is consistent and reliable, with minimal variation across different runs.

The quantum switch circuit shows a rise in entanglement up to three layers, peaking with a concurrence of approximately 0.6, but the trend starts to level off afterward. This suggests that indefinite causal order enhances entanglement but reaches a saturation point beyond which additional layers offer diminishing returns. Interestingly, the standard deviation for the quantum switch starts high but decreases as layers increase, indicating that while early layers lead to more variability in entanglement, deeper circuits stabilize this effect. This characteristic is usually seen in random circuits and circuits with high expressibility. Overall, the initial variability in the quantum switch might be attributed to the complexity introduced by the indefinite order.

In contrast, the time flip circuit exhibits a more modest and gradually increasing entanglement capacity, stabilizing at a concurrence of around 0.5. Its performance is weaker compared to the quantum switch and definite circuits like $R_{xy}$, suggesting that indefinite time-ordering does not particularly provide a strong advantage in generating entanglement. The standard deviation for the time flip circuit shows a gradual decline, indicating an improvement in consistency across layers, but the overall concurrence remains low. This suggests that while the circuit becomes more stable, its entangling potential remains limited.

The $R_{xy}R_{yx}$ circuit presents a different story, with a noticeable decline in concurrence as more layers are added. Starting at around 0.5, its performance drops to 0.4 by the fourth layer. The increasing standard deviation for this circuit, especially in the first two layers, indicates that both controlled rotations lead to more variability in the generated entanglement. As the circuit depth increases, both the entanglement and consistency worsen, suggesting that the complexity introduced by the alternating rotations may interfere with the circuit's ability to generate stable and high levels of entanglement.

Finally, the hardware-efficient ansatz circuit demonstrates a steadily increasing entanglement measure. Its concurrence hovers around 0.4, similar to the time flip circuit, and remains largely unchanged as the number of layers grows after the first one. The high but stable standard deviation reflects the consistency of the HEA circuit, indicating that while it may not excel in generating entanglement, its performance is at least predictable as the number of layers increases. This result suggests that despite the presence of entangling gates, the overall design of the HEA circuit is not optimized for maximizing concurrence, potentially due to its focus on hardware efficiency rather than entanglement generation. This, however, might be from its limitation to only two qubits and two parameters. By expanding the system to more qubits, one might choose different connectivity topologies that better fit their needs with relation to entanglement generation~\cite{correr2024characterizing,correr2024optimal}.

In summary, the $R_{xy}$ circuit not only produces the highest concurrence but also does so consistently with minimal variation. Indefinite circuits, like the quantum switch and the time flip, offer intermediate to lower performance compared to circuits like the HEA and the $R_{xy}R_{yx}$, with initial variability that stabilizes as more layers are added. Based on the concurrence results, we have not found reasons to believe that using an indefinite circuit might provide an advantage in terms of entanglement generation compared to definite ones.

The analysis of the entangling power reveals a distinct trend across different quantum circuits as the number of layers increases. Initially, circuits such as the quantum switch and time flip exhibit lower entangling power compared to the HEA, which starts with the maximum entangling power in the first layer. However, as more layers are added, we observe a convergence in the performance of the indefinite and definite circuits. Specifically, the quantum switch and time flip circuits progressively enhance their entangling power, reaching HEA in deeper layers, as seen at layer 4.

This progression demonstrates that while HEA begins with strong entangling capabilities, indefinite circuits like quantum switch and time flip are capable of increasing their entanglement capabilities as layers increase, with the former maintaining consistently high values after layer 2. The circuits with definite structures, such as the $R_{xy}$ and $R_{xy}R_{yx}$ configurations, show little variation across layers, with entangling power stabilizing around 0.7, highlighting the limitations of some definite circuit structures in terms of growth in entangling capabilities over deeper layers.

The standard deviation analysis, as depicted in the second plot, provides additional nuance to this discussion. The HEA circuit, which initially exhibits the highest entangling power, is accompanied by no standard deviation in the first layer. This distinct behavior comes from the definition of the entangling power itself, where the CNOT operation generates the highest possible entanglement measure for two qubits and since we are taking the average over the initial product states, the rotation operations for the first layer are irrelevant and we are left only with the entangling gate. However, as more layers are added, this standard deviation increases sharply, following a decrease in absolute entangling power, since by adding more rotation and CNOT gates the entanglement previously generated might get destroyed. On the other hand, the indefinite circuits such as quantum switch and time flip show a more stable behavior, with moderate yet consistently lower standard deviations across layers. This stability, especially in the later layers, implies that these circuits while starting with lower entangling power, can maintain consistent entangling performance as they evolve.

When comparing these findings to the concurrence results, we see some disparities that may be due to how entanglement is quantified in each of these measures. While, using concurrence, the $R_{xy}$ circuit has a better generation of entanglement and lower standard deviation than the other circuits, we see better results for the HEA and indefinite circuits when considering the entangling power, also with the better standard deviations as the number of layers increases.

Perhaps the main distinction between those two measures is their relation with the initial state. While the concurrence has a fixed initial state, we must take the Haar average over the space of product states by the definition of entangling power. This could lead us to believe that the entangling power is the most reliable measure and that indefinite circuits could indeed bring some advantage with more layers, however, the initialization of the states is not always arbitrary and might lead to some issues in the context of parameterized circuits and variational algorithms~\cite{larocca2024review}.

\subsection{Quantum average causal effect}

\begin{figure}[t]
	\centering
	\includegraphics[width=\columnwidth]{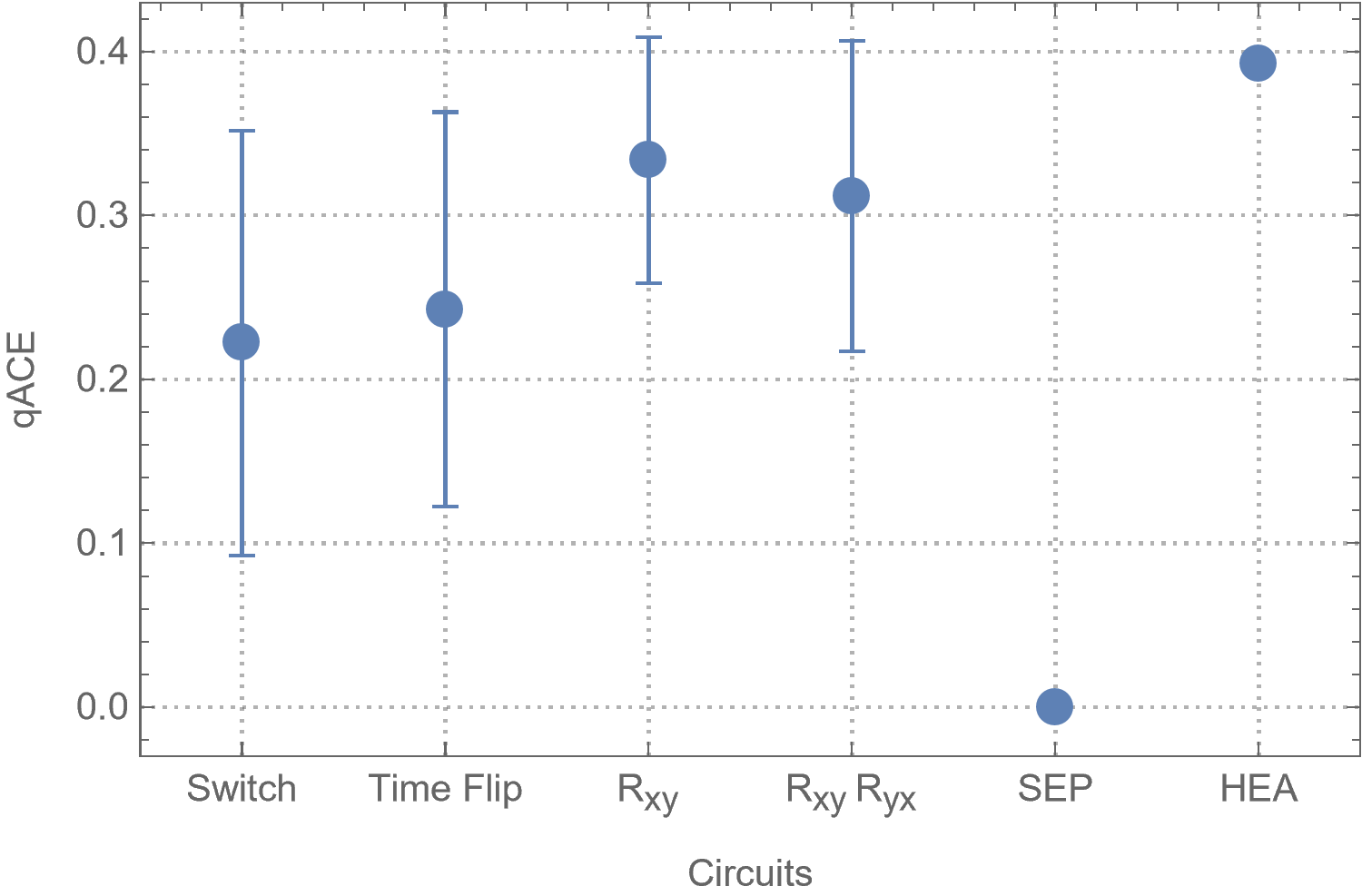}
	\caption{Quantum average causal effect by circuits. The point represents the average value found for different rotational parameters and the bars show their standard deviations. This measure is bounded by zero by separable circuits and by one by the SWAP operation. There is no standard deviation for the last two circuits since one does not have any connections between the first and second qubits (SEP) and the other has this connection unchanged by the choice of parameters (HEA). We also observe that the average for the indefinite circuits is lower than that of the definite ones, while the standard deviation is higher for the former and lower for the latter.}
	\label{qace}
\end{figure}

Lastly, we have the quantum average causal effect measure in Fig.~\ref{qace}. As a reminder, it quantifies the influence of the control qubit input on the output of the target qubit. The circuits that have no connection between the control and target qubits, such as the separable, should measure no influence between each of them.

As expected, in Fig.~\ref{qace} we find the separable circuit with zero qACE values. Another already expected value was the HEA one, with its connection coming from the CNOT gate, already measured in~\cite{hutter2023quantifying} to be $\pi/8 \approx 0.39$.

For the rotationally dependent circuits, we find that the averages for the $R_{xy}$ and  $R_{xy}R_{yx}$ are higher than the ones from the quantum switch and the time flip circuits. This relation between the definite and the indefinite circuits could be explained by previous entanglement results, more specifically, the entangling power for a single layer. In~\cite{hutter2023quantifying} the authors show a correlation between the qACE and the concurrence, however, they have considered uniformly-sampled pure states while we started with the state $\ket{0}$ and the combination of rotations did not lead to a uniform distribution of states in the state space, as we could see with the expressibility result. On the other hand, we have uniformly sampled the initial states to find the entangling power results, and we observe this correlation once more for a single layer, where a higher entangling power corresponds to a higher qACE.

Another remark we can make about this result is that the standard deviations from the indefinite circuits are slightly bigger than the definite ones, indicating a broader range of potential causal interactions between the control and target qubit. Since the number of rotations and parameters is kept constant throughout the tested circuits, this broader range of causal effects could be coming from the indefiniteness of the circuits composed from the quantum switch and the time flip operations.

\section{\label{sec:conclusion}Conclusions}

NISQ computers offer exciting possibilities for various fields, but their noisy nature requires creative solutions. Parameterized quantum circuits and their associated metrics of expressibility and entanglement provide a path forward, while innovative tools like the quantum switch open new avenues for research and practical quantum computing applications.

In this study, we explored the expressibility, entangling power, and quantum average causal effect of parameterized quantum circuits with both definite and indefinite causal structures. We aimed to evaluate how these properties differ across circuit types, particularly with the inclusion of operations such as the quantum switch and quantum time flip, which introduce indefinite causal and temporal order, respectively.

Our findings reveal that circuits with indefinite causal structures, such as the quantum switch, generally exhibit higher expressibility compared to definite circuits, particularly as the number of layers increases. This superior performance suggests that indefinite causal structures enable better exploration of the state space in quantum computations with the same number of adjustable parameters, thus a fair comparison. The time flip circuit, while initially performing on par with definite circuits, eventually approaches the quantum switch’s expressibility with deeper layers.

With regards to the concurrence, definite circuits seem to exhibit high entanglement generation with minimal variation, whereas indefinite circuits display intermediate entanglement performance, with the quantum switch showing initial gains that eventually plateau, and the time flip circuit demonstrating lower but steadily increasing entanglement. These findings suggest that while indefinite causal structures can enhance entanglement, their advantages may diminish as circuit depth grows, and their overall entanglement potential remains lower compared to certain definite circuits.

In terms of entangling power, the study reveals that while certain definite circuits maintain stable entangling capabilities across layers, indefinite circuits like the quantum switch and time flip demonstrate a gradual improvement, particularly in deeper circuits, ultimately converging with or surpassing some definite circuits. This suggests that while definite circuits might provide reliable entangling performance from the outset, indefinite circuits offer the potential for greater entanglement generation with additional layers. Overall, given the disparity between concurrence and entangling power, these results indicate that indefinite circuits may offer advantages in certain regimes, particularly as circuit depth increases, though their performance depends heavily on how entanglement is measured and the specifics of circuit design and initialization.

Finally, the quantum average causal effect reveals that circuits with indefinite causal order exhibit broader potential for causal influence between qubits, evidenced by their larger standard deviations. We have also found that the previously known correlation with concurrence (with uniformly sampled initial states) might also be extended to the entangling power.

In summary, indefinite causal structures offer significant advantages in terms of expressibility, particularly as circuit depth increases. Although their ability to generate entanglement may not always exceed that of definite causal structures, they remain valuable tools for enhancing the capabilities of noisy intermediate-scale quantum devices. Future research should focus on optimizing these indefinite circuits to fully unlock their potential for quantum algorithms and improve their performance in practical quantum computing applications.

\begin{acknowledgments}
	\noindent We are grateful to Daniel Brod and Ernesto Galvão for their valuable discussions and insightful guidance on the subject. This study was financed in part by the Coordenação de Aperfeiçoamento de Pessoal de Nível Superior – Brasil (CAPES) – Finance Code 001 (P.C.A., A.D. and G.I.C.). 
    A.D. acknowledges the financial support from São Paulo Research Foundation—FAPESP (Grant No. 2024/19054-8).
    G.I.C. acknowledges the financial support of the Research Council of Finland through the Finnish Quantum Flagship project (358878, UH).
	I.M. acknowledges financial support from São Paulo Research Foundation - FAPESP (Grants No. 2022/08786-2 and No. 2023/14488-7).
	P.C.A. acknowledges financial support from Conselho Nacional de Desenvolvimento Científico e Tecnológico - Brazil (CNPq - Grant No. 160851/2021-1).
	A.C. acknowledges a license by the Federal University of Alagoas for a sabbatical at the University of São Paulo, and partial financial support by CNPq (Grant No. 168785/2023-4), Alagoas State Research Agency (FAPEAL) (Grant No. APQ2022021000153), and São Paulo Research Foundation (FAPESP) (Grant No. 2023/03562-1).
	D.O.S.P. acknowledges the support by the Brazilian funding agencies CNPq (Grants No. 304891/2022-3 and No. 402074/2023-8), FAPESP (Grant No. 2017/03727-0) and the Brazilian National Institute of Science and Technology of Quantum Information (INCT/IQ).
\end{acknowledgments}

\bibliography{refs.bib,refVQA.bib,refICO.bib}
\bibliographystyle{apsrev4-2}

\end{document}